\documentclass[aps,prx, superscriptaddress,footinbib]{revtex4-2}
\usepackage[makeroom]{cancel}
\usepackage{mathrsfs}
\usepackage{color}
\usepackage{graphicx}
\usepackage{subcaption}
\usepackage[countmax]{subfloat}
\usepackage[english]{babel}

\usepackage{braket}
 \usepackage{dsfont}
\begin{document}
\title{RNA folding landscapes from explicit solvent all-atom simulations}
\author{Gianmarco Lazzeri}
\affiliation{Frankfurt Institute for Advanced Studies, Ruth-Moufang-Straße 1, Frankfurt am Main, D-60438, Germany}
\affiliation{Physics Department of Trento University, Via Sommarive 14, 38123 Povo (Trento), Italy.}
\author{Cristian Micheletti}
\email{cristian.micheletti@sissa.it}
\affiliation{SISSA - International School for Advanced Studies, Via Bonomea 265, 34136 Trieste, Italy.}
\author{Samuela Pasquali}
\email{samuela.pasquali@u-paris.fr}
\affiliation{Laboratoire Biologie Fonctionnelle et Adaptative, CNRS UMR 8251, Inserm ERL U1133,  Université Paris Cité, 35 rue Hélène Brion, Paris, France.}
\author{Pietro Faccioli}
\email{pietro.faccioli@unitn.it}
\affiliation{Laboratoire Biologie Fonctionnelle et Adaptative, CNRS UMR 8251, Inserm ERL U1133,  Université Paris Cité, 35 rue Hélène Brion, Paris, France.}
\affiliation{INFN-TIFPA, Via Sommarive 14, 38123 Povo (Trento), Italy.}
\begin{abstract}
Atomically detailed simulations of RNA folding have proven very challenging in view of the difficulties of developing realistic force fields and the intrinsic computational complexity of sampling rare conformational transitions.
To tackle both these issues, we extend to RNA  an enhanced path sampling method previously successfully applied to proteins. In this scheme, the information about the RNA's native structure is harnessed by a soft history-dependent biasing force, which is added to the atomistic force field, thus promoting the generation of productive folding trajectories.
Here, we report on the results of simulations in explicit solvent of RNA molecules from 20 to 47 nucleotides long and increasing topological complexity.
From a statistical analysis of the folding pathways we infer that, differently from proteins, the underlying free energy landscape is significantly frustrated, even for relatively small chains with a simple topology.
The folding mechanism   and the thermodynamics are in  agreement with the available experiments and some of the existing coarse-grained models.  This scheme provides a fully microscopic characterization of RNA folding, relating the kinetics and dynamics of the transition to the chemistry of the chain and its solvent. Therefore, it provides a transferable framework that sets the stage for future translational applications.
\end{abstract}

\maketitle

 The importance of non-coding RNA molecules has become more and more evident in recent years with the discovery of the central role of these systems in regulating gene expression \cite{Sharp2009,Jiao2014} and other vital cellular processes \cite{Michelini2018}. Moreover, many viruses rely on RNA systems to hijack the host cellular machinery and spread the infection \cite{Balvay2007,Jaafar2019}. Just like for most proteins, to properly function these molecules need to adopt a well defined three-dimensional structure. Understanding their folding can shed light onto their function and may inspire new therapeutic strategies.
 However, the limited chemical alphabet of nucleic acid bases and their ability to form both canonical and non-canonical pairings \cite{Leontis2001}, make RNA folding prone to frustration, as alternative overall architectures can be adopted by the same sequence. This is what happens for riboswitches, where the presence of a ligand triggers a full reorganization of the structure \cite{Garst2011}, but also for other systems, now known to adopt alternative structures \cite{Saldi2021}.
Striking in this sense are the experiments on G-quadruplexes ---formed by both DNA and RNA \cite{Kolesnikova2019,Lightfoot2019,Fay2017}--- and on the hairpin of 7SK RNA~\cite{MartinezZapien2017}. Moreover, even single-point mutations and post transcriptional modifications such as methylations can trigger global structural rearrangements~\cite{Roeder2020,Wang2014,Manners2019}.

RNA folding has been tackled before by computational means in various models \cite{Sponer2018}.
Plain molecular dynamics (MD) simulations at atomistic resolution have been performed for specific RNAs of limited size \cite{Miner2016,Kuehrova2016}. However,  most  RNA molecules of biological interest, comprising from a few dozens to a few hundreds nucleotides,  remain out of reach. More recently, atomistic path-sampling methods have been proposed to identify alternative stable RNA structures and study their low-energy interconversion paths \cite{Joseph2017,Roeder2020,Roeder2021}. These appealing, though computationally demanding, methods require an implicit solvent description, because they are based on a geometrical optimization scheme, rather than on integrating equations of motion.
In addition, different coarse-grained models \cite{Cragnolini2015a,Dawson2016,Li2021},  including native centric ones, have been adopted to overcome limitations of the atomistic approaches. These schemes are used to capture the main features of RNA folding of biologically relevant systems. However, these models  do not provide atomically detailed physico-chemical insight. This limitation prevents potential translational applications, such as those recently introduced for targeting proteins~\cite{PPI-FIT}.

In response to these challenges, we  adopt  a variational enhanced path sampling method that enables the simulation of RNA folding within state-of-the-art all-atom force fields in explicit solvent. In such a scheme, first a specific type of biased dynamics called ratchet-and-pawl MD (rMD) \cite{rMD1, rMD2} is employed to efficiently generate a statistically significant number of productive trajectories. In rMD, an auxiliary history-dependent potential depending on a suitable collective variable (CV) is introduced to prevent the chain from backtracking toward the initial state (unfolded, reactant). Conversely, the biasing force is inactive when the system spontaneously progresses toward the final state (folded, product). In the ideal case  in which the CV is the committor function \cite{TPT1},  rMD  simulations yield the correct Boltzmann sampling in the region explored by the transition path ensemble \cite{irMD,irMDVdE}.
 In our simulation of RNA folding, we use a CV that measures the distance in contact map space of the instantaneous configurations from the known crystal native structure. Since this CV is only a proxy of the ideal reaction coordinate, rMD yields approximate sampling of the equilibrium distribution in the transition region. However, it is still possible to keep systematic errors to a minimum by applying the so-called Bias Functional (BF) filtering~\cite{BFA}. In this approach, a variational principle derived from Langevin dynamics is used for scoring the folding trajectories generated by rMD to identify those with the highest probability of occurring in the absence of any biasing force.

In the context of protein folding this scheme has been successfully validated against available plain MD folding trajectories \cite{BFA, Het-s} and several experiments \cite{SerpinFold, IM7, Jacs1, Jacs2}. The accuracy of the predictions  allowed for rationalizing the effect of  point mutations related to protein misfolding~\cite{SerpinFold} and even provided  an innovative scheme to identify small molecules that can interfere with the  folding  process, leading to a new class of protein degraders\cite{PPI-FIT}.

Extending the scope of such folding simulation methods to
RNA  would open numerous perspectives. For instance, it would allow to detect bifurcation points of folding, determining how the sequence folds on a specific state over the possible alternatives, or identify metastable states that can become dominant upon changes of environmental conditions. This would provide invaluable insights to better understand the folding mechanisms of these systems as well as to envision strategies for drug design down the road.
In this study, we report on  the first extension of this simulation scheme to RNA. Besides characterizing the details of the folding process, we also highlight key differences between folding of RNAs and proteins.

\section*{Results}

As a first step, we used our strategy to estimate the free energy landscapes of different RNAs and expose the thermodynamic forces that underpin their folding processes. The comparison is especially apt at revealing universal landscape properties that are specific to RNAs, meaning that they hold across unrelated RNA molecules and yet have no analog in other types of structured biomolecules, such as proteins.

For this comparative endeavour, that to our knowledge has not been pursued before, we selected two RNAs of comparable length and markedly different secondary and tertiary organization. Specifically, we considered: (i) the domain II of the CCHMVD HammerHead Ribozyme (PDBid: 2RPK  and hereby referred to as HHR)\cite{Dufour2009} and (ii) the PK1 pseudoknot of the {\em Aquifex aeolicus} tmRNA  (PDBid: 2G1W )\cite{NoninLecomte2006}. The former consists of 20 nucleotides and has a typical hairpin structure, while the latter is 22 nucleotides long and adopts a  H-pseudoknot configuration where two stems, of 4 and 3 base pairs respectively, are interlaced.

\begin{figure}
\centering
\includegraphics[width=0.5\linewidth]{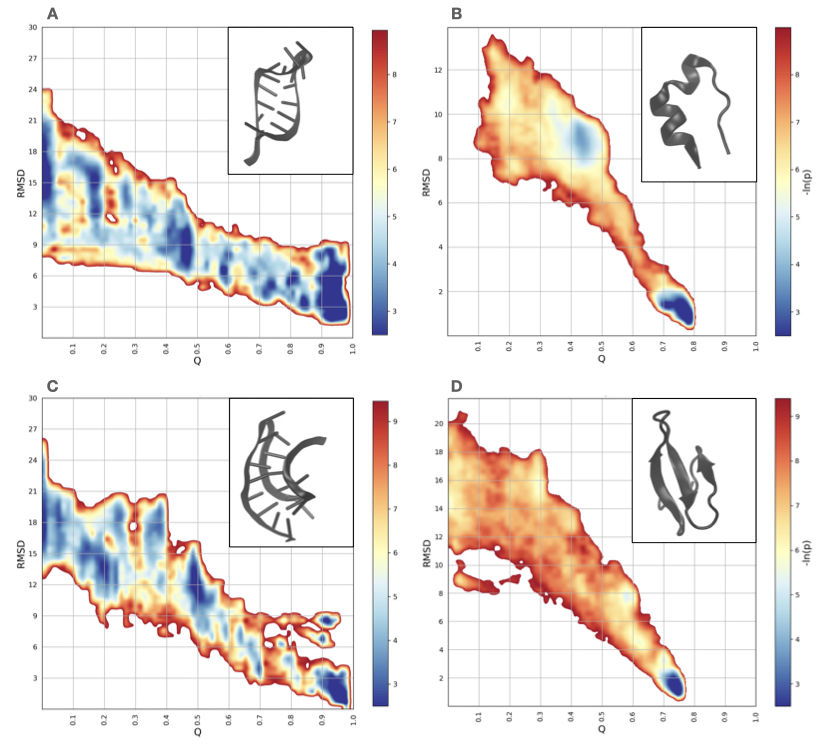}
\caption{Estimate of the free-energy landscape as a function of the fraction of native contacts $Q$ and the RMSD to the crystal native structures for four different macromolecules:(A) HHR, (B) protein TRP-CGE, (C) tmRNA fragement, and (D) protein WW-Domain.  The  landscapes were obtained from a frequency histogram of the successful trajectories obtained by rMD, projected onto the plane of the two collective variables.}
\label{fig:EL RNA-prot}
\end{figure}

{\bf Free energy folding landscape.} The  free energy landscapes of the two molecules estimated from our  simulations are represented by the heatmaps of Fig.~\ref{fig:EL RNA-prot}. The maps were obtained from a frequency histogram of the all-atom folding trajectories projected onto a plane defined by two conventional order parameters, namely the fraction of native contacts $Q$ and the root-mean-squared deviation (RMSD) from the experimental native structure.

The landscapes present a few notable properties. First, they are riddled with local minima with a typical depth of few K$_\textrm{B}$T that can  trap the chain and hinder the folding process. It should be stressed that rMD simulations provide a lower-bound estimate of free energy barriers. Indeed, in an ideal rMD simulation, the biasing force would be proportional to the gradient of the committor function  (see Eq. \ref{rMDF}), which is exponentially small inside metastable states. Therefore, in an ideal simulation, escaping metastable minima becomes exponentially hard, as expected since the trajectories ultimately sample the correct Boltzmann distribution. In realistic conditions, however, the biasing CV  remains finite inside free energy basins, thus enhancing the rate of barrier crossing events.
A second notable feature of our landscapes is that the RMSD-width of the explored region decreases only moderately with growing native-state similarity: As folding progresses, the breadth of the visited conformational space remains substantial, suggesting that the underlying energy landscape is not strongly funneled. It is also interesting to note that the energy minima in the Q-RMSD plane have an oblong vertical structure. This  indicates  a relatively large conformational variability within states of given fraction of native contacts. The emerging picture is one in which once a base pair comes in contact, the system undergoes local structural  adjustments to reach  the right geometric conformation required for pairing. Overall, the landscapes of these two representative RNAs are unexpectedly similar, which is striking given their very different topologies.

This conclusion can be further supported by contrasting the landscapes of HHR  and the tmRNA with those of two protein counterparts, see Fig.~\ref{fig:EL RNA-prot}. For an equal footing comparison, we considered the TRP-CAGE miniprotein (PDB code: 2JOF) and a WW-domain (PDB code:  2F21), which have about the same number of monomeric units and  similar pattern of local vs non-local contacts as the two representative RNAs: The TRP-CAGE consists of 20 amino acids and has a helix structure, while the WW-domain comprises 36 amino acids in a $\beta$-sheet motif.
The comparative inspection of left- and right- panels of Fig.~\ref{fig:EL RNA-prot} shows that both protein counterparts present a smoother landscape with a tapering RMSD-width as the native state is approached. In addition, the metastable minima in the protein  landscapes do not display the oblong shape  observed in the RNAs. Furthermore, the overall slope of the  landscape in the RMSD-Q plane is steeper for proteins than for RNA.  These features coherently indicate that folding is more cooperative for proteins, in agreement with the minimal frustration principle~\cite{Dill1997}. This underscores the much higher complexity of the free-energy landscape of RNAs.

{\bf Folding kinetics.} The several minima that line the free-energy landscape of the two representative RNAs at various stages of the folding process suggest that the native state similarity, or native overlap, is likely not the only salient reaction coordinate for the process. Indeed, the variety of minima are indicative of states that hinder the progressive build-up of native contacts.

\begin{figure}[t]
\centering
\includegraphics[width=0.5\linewidth]{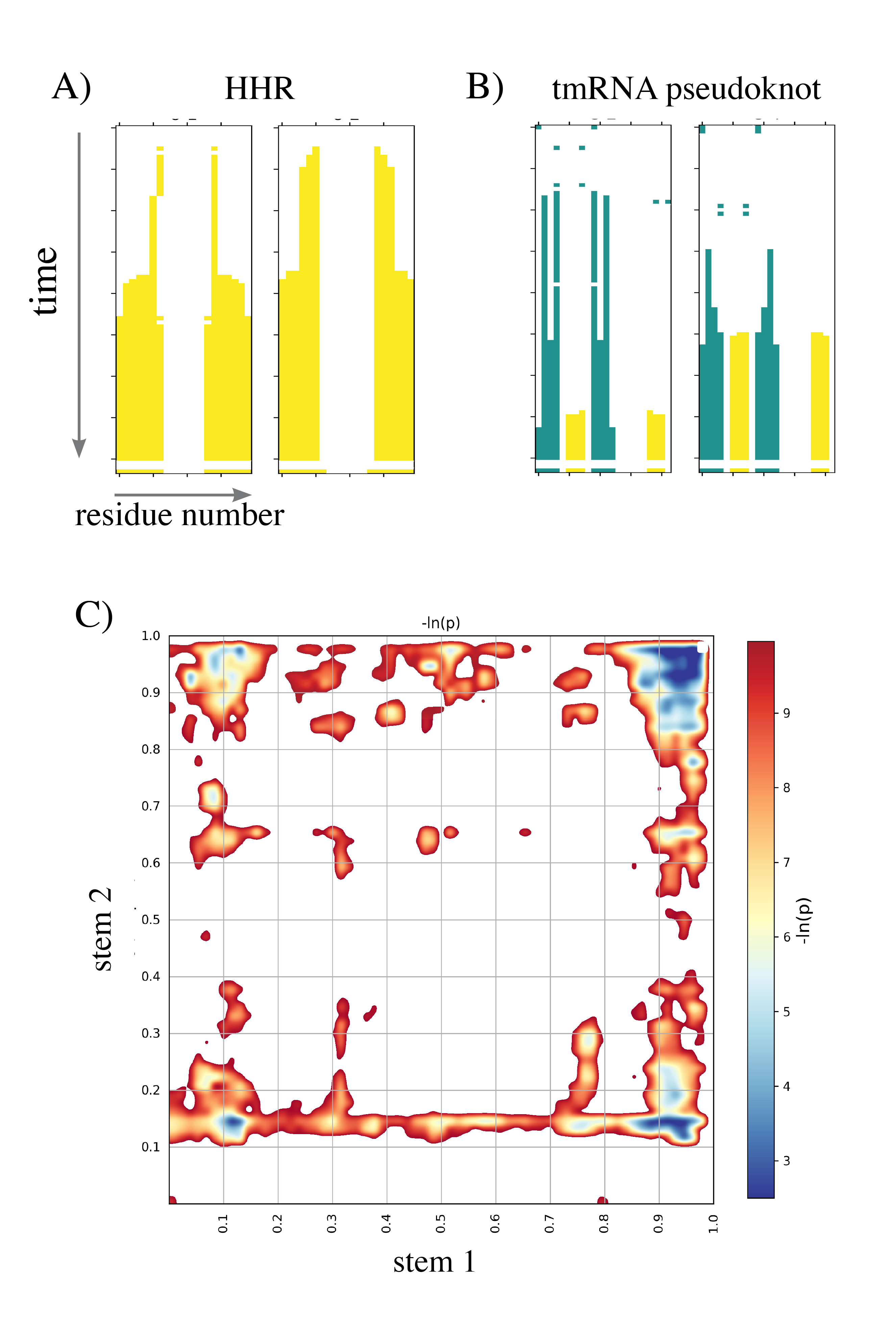}
\caption{Examples of secondary structure formation in folding trajectories for the simple hairpin (A) and for the H-pseudoknot (B). In each graph the vertical axis indicates time from top to bottom while residue number is on the horizontal axis. At each time a base pair is formed it is colored according to a color code identifying the stem (HHR: yellow, tmRNA pseudoknot: stem 1 in green, stem 2 in yellow). B) Free-energy landscape as a function of the fraction of stem formed for tmRNA pseudoknot. }\label{fig:order}
\end{figure}

We observed that the formation of the stems is a key step of the folding process for both representative RNAs, albeit with important differences informed by their architectures, see Fig.~\ref{fig:order}.
For HHR, the most explored folding pathway is one in which the transition is nucleated by the formation of key contacts at the turn of the loop, following a "zipping mechanism". This is exemplified by the two folding trajectories of  Fig.~\ref{fig:order}A. Zipping from the loop toward the extremities can proceed more or less rapidly and is by far the dominant mechanism: Among all  trajectories that reached the native structure (49 out of 100) only one exhibited an inverse zipping mechanism, i.e.~with the ladder of pairings starting from the termini.

In contrast, the tmRNA fragment preferentially folded by first forming the 4 base pairs of stem 1 (G1-G4:C10-C13) followed by the base pairs of stem  2 (G6-C8:G19-C21), as illustrated in Fig.~\ref{fig:order}B. These plots further show that in the early folding stages,  native base pairs can form and then break, in spite of the presence of the rMD bias that promotes a growing overlap with native contacts.
As shown by the typical example of base-pair formation reported in Figs.~\ref{fig:order}A and \ref{fig:order}B, the early stage of RNA folding does not involve the formation of base pairs alternative to  natives ones.

The preferential route of folding stem 1 followed by stem 2 can also be observed from the overall population distribution as a function of the percentage of formation of each stem (Fig.~\ref{fig:order}C). Notably,  stem 1, which is the first to fold,  has a larger number of GC pairs (4) than stem 2 (3), in line with the  correlation between folding order and thermodynamics stability reported in Refs.~\cite{Tinoco99,Cho2009}.

\begin{figure*}[t]
\centering
\includegraphics[width=0.5\linewidth]{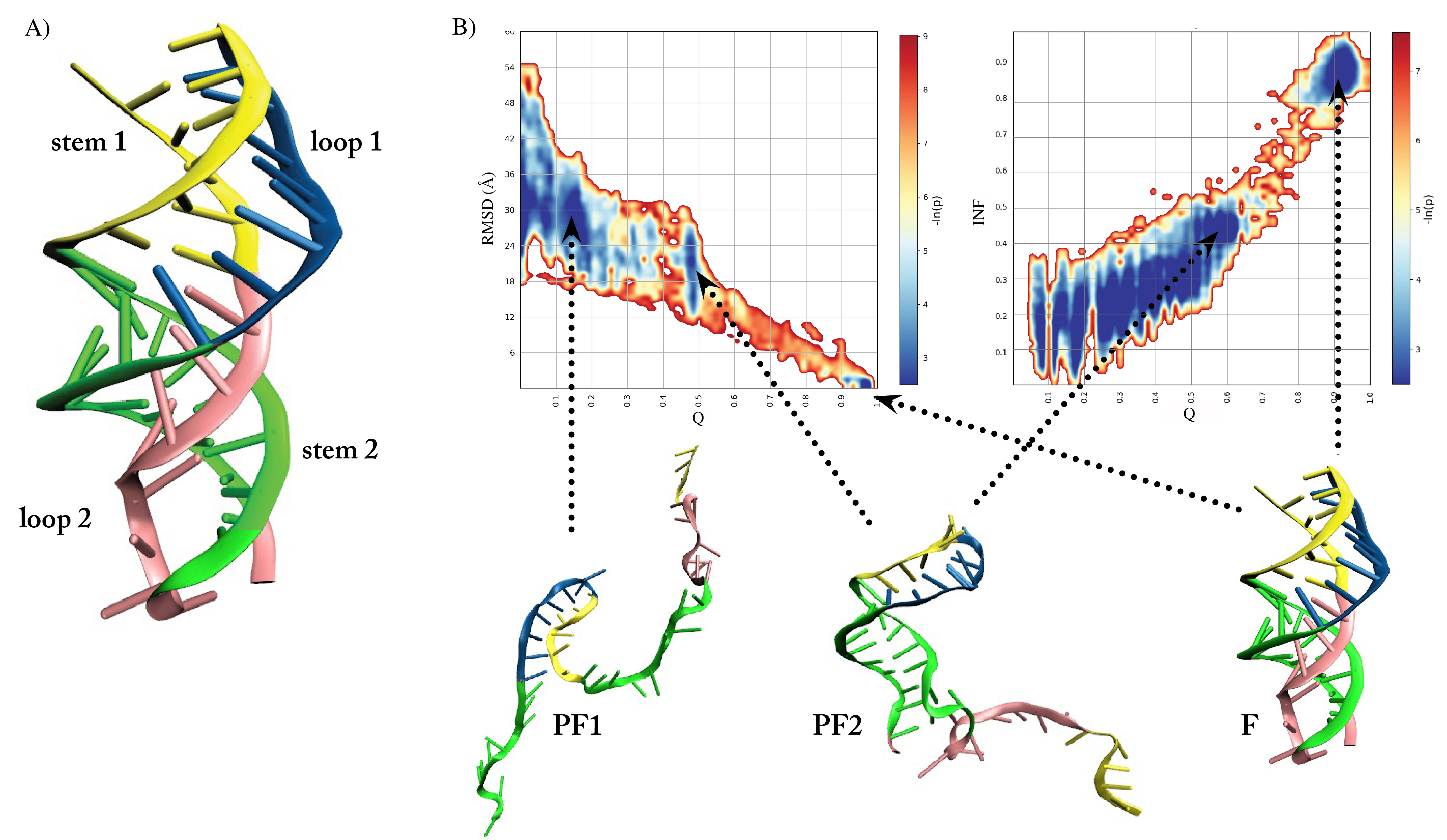}
\caption{A: Native structure of hTR with stems and loops highlighted.
B top: estimate of the free-energy landscape as a function of RMSD (left) and Interaction Network Fidelity - INF - (right) and as a function of the fraction of native contacts $Q$ for hTR. B bottom: examples of structures belonging to the main basins identified by the landscapes. Left: one of many partially unfolded structures with loop 1 folded (PF1); center: stable partially unfolded structure with loop 1 and stem 2 formed (PF2); right: fully folded triple helix pseudoknot (F).}\label{fig:2k96_EL}
\end{figure*}

Having discussed the qualitative features that can be inferred from these simulations, we now address the problem of accessing their accuracy against the available experimental results.
To this goal, we simulated the folding of the P2B-P3 pseudoknot from human telomerase (delta U177 variant, PDB id: 2K96, referred to as hTR) \cite{Kim2008}, which has been extensively characterized by NMR, calorimetry and FRET experiments \cite{Theimer2005, Gavory2006}, and also studied using different coarse-grained models \cite{Cho2009,Denesyuk2011, Biyun2011,  Cragnolini2015}.
This molecule consists of 47 nucleotides and in its native state adopts the conformation of a H-pseudoknot with two stems and two loops. The longer loop (loop 1) wraps around one of the stems (stem 2) forming a series of successive U-A-U triplets and a triple helix. This gives the molecule a straight and compact shape (Fig.~\ref{fig:2k96_EL} A).

As for the previous systems, we looked at the free-energy landscape as a function of fraction of native contacts $Q$ and the native RMSD (Fig.~\ref{fig:2k96_EL}). As for the previous representative RNAs, we note that the energy landscape is very rugged. However, in this case, there is a more pronounced bottleneck near the native state.

We observe a first basin populated along the folding pathway that features loop 1 formed (PF1 in Fig.~\ref{fig:2k96_EL}). At a later stage, the system populates a second basin, where also stem 2 is formed (PF2). Subsequently, the triple helix forms (loop 2) and
 stem1 is established, yielding the native state (F).

For complex RNA architectures,  the RMSD CV might  not be sufficient  to characterize the folding process \cite{Miao2017,Magnus2020}. For this reason, we additionally considered  the interaction network fidelity (INF) parameter \cite{Parisien2009} computed on the number native and non-native base pairs.

\begin{figure}[t]
\centering
\includegraphics[width=0.5\linewidth]{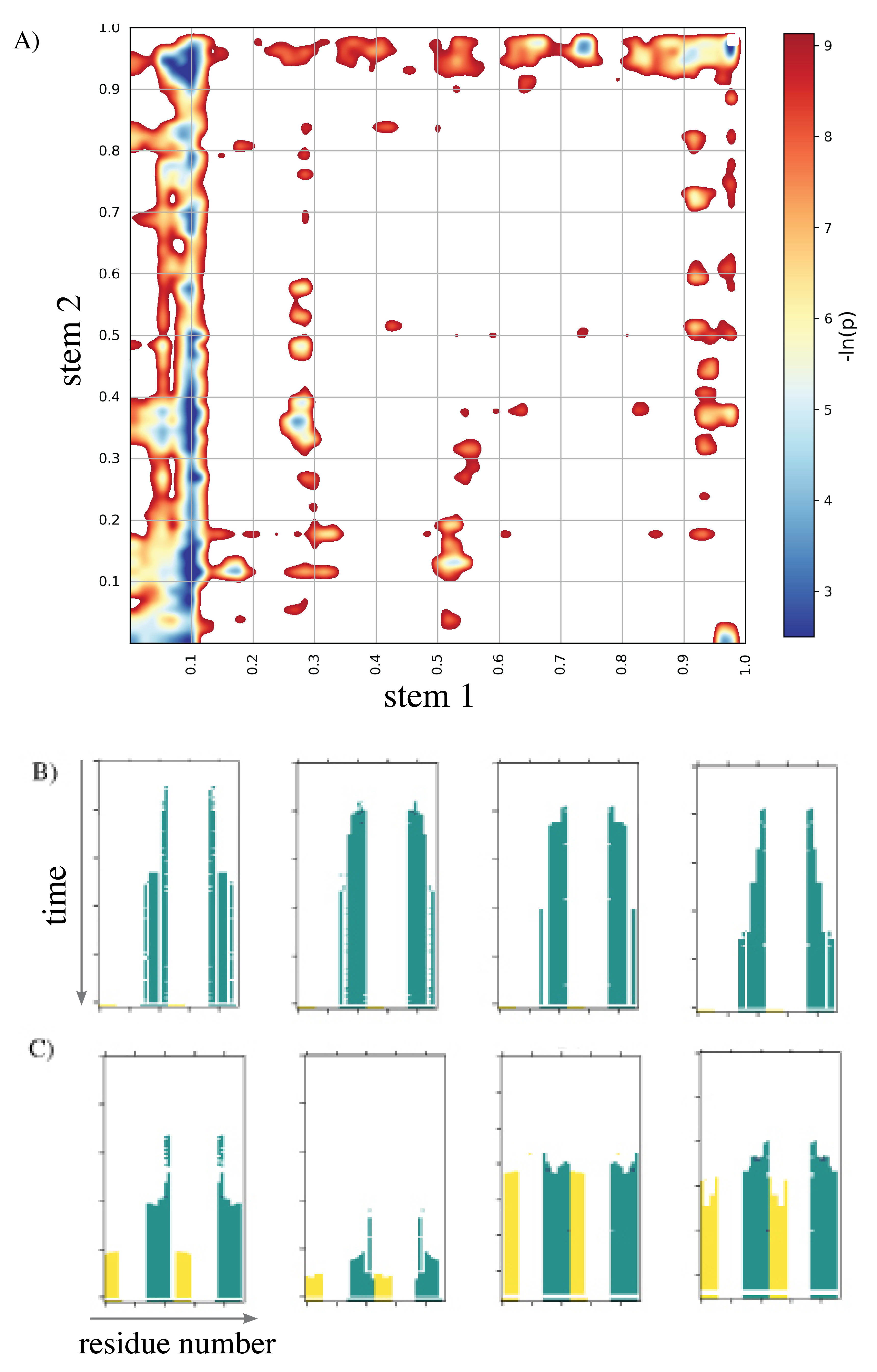}
\caption{A: free-energy landscape as a function of the fraction of folding of stem 1 and stem 2.
Examples of secondary structure formation in rMD trajectories of 2K96 triple helix leading to partially folded structures (B) and to  trajectories leading to complete folding (C). Stem 1 is highlighted in yellow and stem 2 in green.}\label{fig:2k96_2D}
\end{figure}

When we analyze separately the formation of the two stems composing the pseudoknot we see that in the main route to folding  stem 2 forms first (Fig.~\ref{fig:2k96_2D} A). As in the previous case,   this stem folds through a zipping mechanism. In contrast, the shorter stem 1 folds more cooperatively (Fig.~\ref{fig:2k96_2D} B and C).
This folding mechanism is in qualitative agreement with the results of thermal denaturation experiments \cite{Theimer2005}.
As also discussed in previous simulation studies of the same system~\cite{Cho2009}, from the observation of the hierarchical folding of RNA molecules~\cite{Li2008} it is possible to draw a connection between thermal stability and order of events in folding.
Thermal denaturation of  the $\Delta U177$ mutant of hTR occurs at a melting temperature of around $70^\circ$C for both stems. However, melting of the loops differs significantly, with $T_m \sim 60^\circ$ for loop 1 and  $T_m \sim 70^\circ$C for loop 2.
This hints at a faster folding of loop 2 than loop 1.
Moreover, the interactions forming loop 1 are more short-ranged than those of the two stems, which have comparable enthalpy. Therefore, we expect loop 1 to be the initial folding step.
The observation that folding of loop 1 leads to folding of stem 2, suggest the earlier folding of stem 2  compared to stem 1.
Our landscape agrees with this thermodynamic analysis, with loop 1 being the first structure to form, followed by stem 2, and then stem 1.
This preferred pathway has also been observed in coarse-grained simulations of the same system, both with native-driven~\cite{Cho2009} and with unbiased simulations~\cite{Cragnolini2015}.

\section*{Discussion}

Our reconstruction  of folding landscapes of RNAs and proteins of similar length and topology highlights key differences between  these two classes of biopolymers.
In particular, the energy landscape of RNAs is significantly more frustrated, displaying a larger number of local minima, and overall less funneled.
This is suggestive of a larger conformational heterogeneity of the RNA equilibrium ensembles. These results, along with those of other recent path sampling simulations \cite{Joseph2017, Souza2017,Cragnolini2017,Roeder2020,Roeder2021}, structure prediction \cite{Schroeder2018, Denesyuk2011, Cragnolini2015, Cruz2012,Miao2015, Miao2017} and experiments \cite{Ritz2013, Wu2020,Yu2021} coherently point toward the conclusion that RNAs are moderately frustrated systems, with energy landscapes displaying several alternative competing minima for any given sequence.

The computational cost of generating many folding pathways for the hTR pseudo-knot in explicit solvent was relatively low, enabling our simulations to be completed in a just few days on a small GPU cluster. Our calculations are already in good  agreement with thermodynamic experiments and previous simulations performed in coarse-grained models, but do not allow us to obtain information about kinetics. Indeed, our estimates of the relevant energy barriers are, in general, affected by errors induced by suboptimal choices of biasing CVs. In particular, in this first application to RNAs, we chose to adopt a CV that was  inspired by energy lanscape theory arguments that are well applicable to protein folding, but may not be as fit for more frustrated system. This limitation could be overcome by employing a biasing CV obtained with more advanced techniques, e.g. using self-consistent refinement schemes \cite{SCPS, PerezCV} or machine learning methods \cite{TuckCV, CovinoCV, BolhuisCV, ParrinelloCV}.

From the observed agreement  with experimental data and the high computational efficiency, we envision that the general scheme underlying our method could be used to explore the folding of broad range of RNAs  with fully atomistic physico-chemical detail.
Possible applicative avenues include the characterization of  the folding pathways to alternative structures, such as those found in G-quadruplexes, riboswitches, and pseudoknotted motifs of functional relevance\cite{Suma2020,Schlick2021,Schlick2021b}. Using the same method it would also be possibile to address the question of what are the key branching points in the folding pathways associated to different native RNA structures and gain  insight into the underlying physico-chemical mechanisms.

\section*{Materials and Methods}

\subsection{RNA folding Simulation}
RNA folding simulations were carried out using the AMBER ff14SB force field \cite{AMBER99SB} in explicit TIP3P water. After imposing charge neutrality on the system by introducing monovalent sodium ions, we further added a buffer of  0.15 M sodium chloride. The size of the cubic periodic simulation box was chosen large enough to  accommodate  the  fully unfolded conformations. The total number of atoms in our systems ranged from tens of thousands (for the two  smallest RNAs) to several hundreds of thousands (for the largest RNA).

For each of these three nucleic acids, we sampled their native states with 100 ns of plain MD at room temperature (310 K) and standard pressure starting from the energy-minimized PDB configuration and discarding the first 10 ns. To keep  temperature and pressure fixed, we employed a stochastic velocity re-scaling procedure~\cite{Bussi2007}.
We used this local equilibrium sampling to compute the native contact maps $C^0_{ij}$ that are used to define the CV that is employed in rMD simulations (see below).

To obtain 10 initial unfolded configurations for the hairpin and PK1 pseudoknot, we performed MD runs at the nominal temperature of 800K, starting from configurations sampled in the native state.
For hTR, which is more compact,  we instead relied on UnityMol \cite{UnityMol}, a software enabling the interactive visual manipulation of coarse-grained RNAs, to generate efficiently 3 fully unfolded structures. We relaxed all the unfolded RNA structures with 5 ns of MD at standard temperature and pressure  conditions. Then, we initiated a set rMD simulations from each unfolded configuration obtained at the end of this relaxation run. In particular,  10 independent rMD trajectories for each configuration of  the hairpin and the PK1 pseudoknot, and 20 for each configuration of the hTR pseudoknot. The protocol has been previously employed in protein folding simulations, as detailed, e.g., in Ref.~\cite{Het-s}.

In rMD, the history-dependent biasing force is defined as:
 \begin{eqnarray}\label{rMDF}
 {\bf F}_i(X, \tilde{q}_m(t) ) = - k_\textrm{rMD} \nabla_i \tilde{q}(X) (\tilde{q}(X)-\tilde{q}_\textrm{m}(t)) \theta[\tilde{q}(X)-\tilde{q}_m(t)].\nonumber\\
\end{eqnarray}
 In this equation, $X=({\bf x}_1, \ldots,{\bf x}_N)$ denotes the set of  atomic coordinates in the nucleic acid, and
 \begin{equation}\label{qdef}
 \tilde{q}(X) = \frac{\sum'_{i>j} (C_{ij}(X)-C^0_{ij})^2}{\sum'_{i,j}C^{0\, 2}_{ij}}
 \end{equation}
 measures the instantaneous overlap with the native contact map.
 The symbol $\sum'$ denotes a summation which excludes atoms that belong to neighboring aminoacids or nucleotides, to weaken the effect of the rMD bias in forming local contacts. $C_{ij}(X)$ is a switching function that approaches 1 when atom $i$ and $j$ are in contact and vanishes when they are far apart:
\begin{eqnarray}
C_{ij}(X) = \frac{1- \left(\frac{r_{ij}}{r_0}\right)^6}{1- \left(\frac{r_{ij}}{r_0}\right)^{10}}.
\end{eqnarray}
Here, $r_{ij}= |{\bf x}_i - {\bf x}_j|$ and $r_0=4.5 $\AA~ is a threshold reference distance and $C_{ij}^0$ is obtained from an average of $C_{ij}(X)$ over the configurations that have been generated by 100 ns of  MD in the native state.
$\tilde{q}_\textrm{m}(t)$ in Eq.(\ref{rMDF}) is the minimum value attained by the collective variable $\tilde{q}[X(t)]$ till time $t$. The bias indirectly promotes the formation of the native contacts encoded in  $C_{ij}(X_0)$ by counteracting any conformational change that decreases the overlap of $C_{ij}(X)$ with $C^0_{ij}$.

The constant $k_\textrm{rMD}$ was set to $5 \times 10^{-3}$~kJ/Mol for the hairpin and the PK1 pseudoknot, and $1 \times 10^{-3}$~kJ/Mol for the hTR pseudoknot. These values are typical for rMD protein folding simulations, where
they allow for efficiently generating many folding trajectories, while applying a gentle total biasing force. Indeed, the biasing force provided only a small correction to the physical force acting on each atom.  In RNA folding simulations, we noticed that the same choice of $k_\textrm{rMD}$ leads to larger biasing forces, which can become comparable with the physical ones. This is in line with the fact that our choice of CV is inspired by energy landscape theory arguments, which have been introduced in the context of protein folding and may be less fit for RNAs, which are more frustrated systems.

To reduce the effect of the bias, we resorted to the so-called Bias Functional method~\cite{BFA}.
According to this scheme, rigorously derived for Langevin dynamics, the rMD trajectories that have the largest probability to occur in an \emph{unbiased} simulation are those with the least value of the f functional
\begin{equation}
    T = \sum_{i} \frac{1}{m_i \gamma_i} \int_0^{t_f} d\tau | {\bf F}^i_B(X,t)|^2,
\end{equation}
where $t_\textrm{f}$ is the time duration of the trajectory and $m_i, \gamma_i$  denote the mass and viscosity coefficient of each atom, respectively.
We used the BF criterion to identify and discard atypical transitions, i.e.~outliers characterized by a very large value of $T$.

The rMD simulations were run for a total of 2 ns (for hairpin and PK1 pseudoknot) and  5 ns (for hTR pseudoknot), but were terminated earlier if the system reached the native state, defined by a RMSD to native $\in [0, 3]\ \AA$ and fraction of native contacts $Q \in [0.8,1]$.

\subsection{Simulation analysis}
Folding trajectories have been analyzed through software specifically designed to detect key interactions in RNA structures.
In particular, we monitored the RMSD to native, the number of heavy-atom contacts (native and non-native), and the number of base-pairings. The latter CV was used to monitor the formation of the secondary structures.
RMSD to native and contacts were computed using the MDtraj python package~\cite{MDtraj}, with a contact distance cutoff of 4.5\AA.

For each structure in the trajectory, we computed base-pairing based on geometric criteria using the DSSR~\cite{Lu2015} software. Since the detection of hydrogen bonds can depend on the specific cutoffs employed by the algorithm, for a sample of structures we computed base pairing also using Barnaba~\cite{Bottaro2019} to verify the robustness of the results with an independent algorithm, and found a substantial agreement between the two methods.
We also monitored  the formation of multiplets, since they appear in the native structure of hTR, and in the triple helix.
An assessment of the quality of RNA secondary structures  inferred from base-pairing in a sampled conformer was obtained using the Interaction Network Fidelity~\cite{INF}, defined as:
\begin{equation}
\mathrm{INF} = \sqrt{\mathrm{Specificity} \times \mathrm{Sensitivity}},
\end{equation}
where
\begin{eqnarray}
\mathrm{Specificity} = \frac{|TP|}{|TP|+|FP|} \nonumber \\
\mathrm{Sensitivity} = \frac{|TP|}{|TP|+|FN|} . \nonumber
\end{eqnarray}
In these equations, $TP$ is the number of  true positives, i.e. the number of native base pairings, $FP$ the number of false positive, i.e. the number of non-native base pairings, and $FN$ are the false negatives, i.e.~the number of native base pairings not present in the sampled conformer.

\end{document}